\documentclass[12pt]{iopart}

\usepackage{epsfig}
\usepackage{colordvi}
 \newcommand{\IGN}[1]{{}}

\begin{document}
\mbox{}\hfill{New Journal of Physics, in print (2008)}\\

\title{Floquet Stability Analysis of Ott-Grebogi-Yorke and Difference Control}


\author{Jens Christian Claussen}
\address{Institut f\"ur Theoretische Physik und Astrophysik,
Christian-Albrechts-Universit\"at, 24098 Kiel, Germany}
\date{June 4, 2007}


\begin{abstract}
Stabilization of instable periodic orbits of
nonlinear dynamical systems has been a widely explored
field theoretically and in applications.
The techniques can be grouped in time-continuous
control schemes based on Pyragas, and the two
Poincar\'e-based chaos control schemes, Ott-Gebogi-Yorke (OGY) and 
difference control.
Here a new stability analysis 
of these two Poincar\'e-based chaos
control schemes is given by means of Floquet
theory. 
This approach allows to calculate exactly the stability restrictions
occuring for small measurement delays and for an impulse length shorter than
the length of the orbit.
This is of practical experimental relevance; to avoid a selection of the
relative impulse length by trial and error, it is advised to investigate
whether the used control scheme itself
shows systematic limitations on
the choice of the impulse length.
To investigate this point, a Floquet analysis is performed.
For OGY control the influence of the impulse
length is marginal.
As an unexpected result, 
difference control fails when the impulse
length is taken longer than
a maximal value that is 
approximately one half of the orbit length
for small Ljapunov numbers and decreases
with the Ljapunov number.
\end{abstract}

\maketitle


\section{Introduction}
Controlling chaos, or stabilization of
instable periodic orbits of chaotic systems,
has matured to a field of large 
interest in theory and experiment
\cite{hubler,Ott90,Pyragas92,Bielawski93a,Pyragas95,%
Kittel95,tracking,Janson,%
boccaletti,%
handbook2,schusterjust,%
baba02,%
halfperiod,%
parmananda99,pyragas01,ahlborn04,fiedler07,Fukuyama02}.
Most of these techniques are applied to dissipative
systems, 
but with respect to applications as 
suppression of transport in plasmas,
there has been recent interest also in 
controlling chaos
in Hamiltonian systems
\cite{C1,C2,C3,C4,C5,C6,C7}
 and in the stability of  
periodic orbits in multidimensional systems 
\cite{C8}.
Control of chaos in general, however, 
does not rely on the existence of a Hamiltonian,
so the system can be given as any dynamical system described by
a set of differential equations, 
or can be an experimental system.
The aim of chaos control is the stabilization of unstable periodic orbits (UPOs),
or modes in spatial systems.
The central idea 
of the H\"ubler and Ott-Grebogi-Yorke approaches 
to the control of chaos
is to utilize the sensitive dependence
on 
initial conditions, 
be it
{\sl system variables}
\cite{hubler},
or 
{\sl system parameters}
 \cite{Ott90}
to steer the trajectory on an
(otherwise unstable) periodic orbit of the system.
To accomplish this task, 
a feedback is applied to the system
at each crossing of 
a suitably chosen Poincar\'e{} plane,
where the feedback is set proportional to the
actual deviation of the desired trajectory.
Contrary to Pyragas control \cite{Pyragas92}, 
where control is calculated quasi continuously at a 
high sampling rate,
here the approach
is to stabilize by a feedback
calculated at each Poincar\'e section,
which reduces the control problem 
to stabilization of an unstable fixed
point of an iterated map.
The feedback can
be chosen proportional to the distance
to the desired fixed point (OGY scheme), 
or proportional to the
{\sl  difference} 
in phase space position
between actual and last but one 
Poincar\'e section.
This latter method, 
{\sl  difference control}
\cite{Bielawski93a},
or Bielawski-Derozier-Glorieux control,
being a time-discrete counterpart of 
the Pyragas approach \cite{Pyragas92},
allows for stabilization of
inaccurately known
fixed points, 
and
can be extended by
a memory term \cite{Claussen98a,Claussen98c}
to
improve stability
and
to allow for tracking 
\cite{tracking}
of drifting fixed points 
\cite{Claussen98a}.

In this paper the stability of perturbations $x(t)$
around an unstable periodic orbit being subject to
a Poincar\'e-based control scheme is analyzed by means
of Floquet theory \cite{Hale93}.
This approach allows to investigate viewpoints that 
have not been accessible by considering only the iteration dynamics 
between the  Poincar\'e sections,
as measurement delays and variable impulse lengths.
The impulse length is, both in OGY and difference control,
usually a fixed (quasi invisible) parameter; 
and the iterated dynamics is uniquely definded only 
as long as this impulse length is not varied.
The 
influence of the 
impulse length
 has not been point of consideration before; if
mentioned at all, usually
a relative length of approximately 1/3 is chosen 
without any reported sensitivity.
Whereas for the Pyragas control method 
(in which the delayed state feedback enforces a time-continuous 
description) a Floquet stability analysis is known \cite{Just97}, 
here the focus is on the time-discrete control.

\subsection{Floquet stability analysis}
The linearized differential equations of both schemes
are invariant under translation in time, $t\to{}t+T$.
Herby, we assume that the system under control is not explicitely
time-dependent.
According to the theory of delay-differential equations 
\cite{Hale93},
a stability condition can be derived from a simple 
eigenvalue analysis of Floquet modes.
The Floquet ansatz expands the solutions
after periodic solutions
 $u(t+T)=u(t)$ 
according to
\begin{eqnarray}
x(t) = {\rm{}e}^{\gamma t} u_\gamma(t).
\end{eqnarray}
The necessary condition 
on the Floquet multiplier ${\rm e}^{\gamma T}$
of an orbit of duration $T$
for stability of the solution
is  Re$\gamma<0$;
and $x(t)\equiv{}0$ refers to motion along the orbit.
In Poincar\'e-based control the effective
motion can be transformed into the unstable eigenspace,
see e.g.\ App.\ A in \cite{Claussen98c};
the stability is governed by the motion therein.
For the case of one unstable Lyapunov
exponent,
 this
subspace is one-dimensional.

\subsection{OGY control}
The method proposed by Ott, Grebogi and Yorke \cite{Ott90}
applies a control amplitude 
$r(t)=\varepsilon (t) (x(t_\times) -x^*)$,
in vicinity of a fixed point $x^*$.
Here $\varepsilon (t)$ is a 
(possibly time-dependent)
feedback gain parameter,
and $x(t_\times)$ is the position of the last Poincare crossing.
Without loss of generality, we can place the fixed point at 
$x=0$, so that the OGY feedback scheme becomes
$r(t)=\varepsilon(t) x(t_\times)$, where 
$t_\times \equiv t-(t ~{\rm{}mod}~ T)$ is the
time of the last Poincare crossing. 
Now one considers the linearized motion in 
vicinity of an unstable periodic orbit
(which is a stable periodic orbit of a successfully 
controlled system).
The Poincar\'e crossing reduces the dimensionality from
$N$ to $N-1$ dimensions, in the lowest-dimensional
case from 3 to 2. 
In this case 
it is sufficient to
consider a linearized 
one-dimensional time-continuous motion
around the orbit,
$\dot{x}(t)=\lambda x(t) + \mu r(t)$,
which now can be complex-valued to account for 
flip motion around the orbit \cite{Just97},
i.e., 
one has the dynamical system
\begin{eqnarray}
\dot{x}(t)=\lambda x(t) + \mu\varepsilon x( t-(t ~{\rm{}mod}~ T)).
\end{eqnarray}
 Without control ($r(t)=0$), the time evolution of this 
 system is simply $x(t)={\rm e}^{\lambda t}$ and the
 Ljapunov exponent of the uncontrolled system is
 ${\rm Re}\lambda$.
Here it must be emphasized that assuming
constant $\lambda$ and $\mu$
is a quite crude approximation, so only 
qualitative results can be concluded.
Now we see 
-- a central observation --
that no 
delay-differential equation
\cite{Hale93}
is obtained:
As the ``delay'' term always refers to the last Poincar\'e 
crossing, 
this type of
 dynamics can be integrated
 piecewise.
In the first time interval between $t=0$ and $t=T$ 
the differential equation reads
\begin{eqnarray}
\forall_{0<t<T} \;\;\;\;\;\;\;\;
\dot{x}(t)=\lambda x(t) + \mu\varepsilon x(0).
\nonumber
\end{eqnarray}
Integration  of this differential equation yields
\begin{eqnarray}
x(t) =
\left((1+  \frac{\mu\varepsilon}{\lambda} )
{\rm{}e}^{\lambda t} - \frac{\mu\varepsilon}{\lambda} 
\right) x(0).
\nonumber
\end{eqnarray}
This gives us an iterated dynamics (here we label the begin of the
time period again with $t$)
\begin{eqnarray}
x(t+T) = 
\left((1+  \frac{\mu\varepsilon}{\lambda} )
{\rm{}e}^{\lambda T} - \frac{\mu\varepsilon}{\lambda}
\right)
x(t).
\nonumber
\end{eqnarray}
This equation allows to determine the Floquet modes
$x(t+T)={\rm e}^{\gamma T} x(t)$ by inspection.
The  Floquet multiplier ${\rm e}^{\gamma T}$ of an orbit,
assuming an impulse duration of full orbit length,
 therefore is given by
\begin{eqnarray}
{\rm{}e}^{\gamma T}= (1+  \frac{\mu\varepsilon}{\lambda})
{\rm{}e}^{\lambda T} - \frac{\mu\varepsilon}{\lambda}.
\end{eqnarray}
The remainder of the paper investigates,
both for OGY and difference control,
how the Floquet multiplier 
is modified for different
impulse lengths.

%
\IGN{
Using a common notation for real and imaginary parts
$\lambda=\lambda'+ {\sf{}i} \lambda''$ 
and $\gamma = \gamma'+ {\sf{}i} \gamma''$,
for fixed values of 
$T$ and $\mu\varepsilon$ 
one can map the whole
 ${\rm{}Re}\gamma=0$  line
into the complex
 $\lambda$ plane.
With this ingredients, the condition
 $\gamma'\stackrel{!}{=}0$ 
reads
\begin{eqnarray}
(\lambda'+ {\sf{}i} \lambda'')
(\cos\gamma''T + {\sf{}i} \sin\gamma''T) 
=
 \mu\varepsilon 
~~~~~~~~~ ~~~~~~~~
~~~~~~~~~~
\nonumber
\\
\nonumber
~~~~~~~~
+( -\mu\varepsilon + \lambda'+ {\sf{}i} \lambda'')
(\cos\lambda''T {\rm{}e}^{\lambda'T} 
 +  {\sf{}i} \sin\lambda''T {\rm{}e}^{\lambda'T} ), 
\nonumber 
\end{eqnarray}
being equivalent to a system of equations
for real and imaginary part
\begin{eqnarray}
\lambda'\cos\gamma''T 
- \lambda''\sin\gamma''T
\nonumber
&=&
\mu\varepsilon
+(- \mu\varepsilon + \lambda') \cos\lambda''T {\rm{}e}^{\lambda'T} 
\nonumber \\ 
\nonumber 
& & 
-\lambda'' \sin\lambda''T {\rm{}e}^{\lambda'T} 
\\
 \lambda''\cos\gamma''T + \lambda' \sin\gamma''T
\nonumber 
&=&
\lambda'' \cos\lambda''T {\rm{}e}^{\lambda'T}
\nonumber \\ 
& & 
+( -\mu\varepsilon + \lambda')  \sin\lambda''T {\rm{}e}^{\lambda'T}.
\end{eqnarray}
This is a system of 
eqns.
 transcendent in both
$\lambda'$ and $\lambda''$.
\\
\noindent
Instead of tackling this system directly for
special cases of the parameters, one 
instead can plot the real part of the Floquet multiplier
as a function of the parameters.
}

\section{Influence of impulse length: OGY case}
\begin{figure}[htbp]
\noindent
\epsfig{file=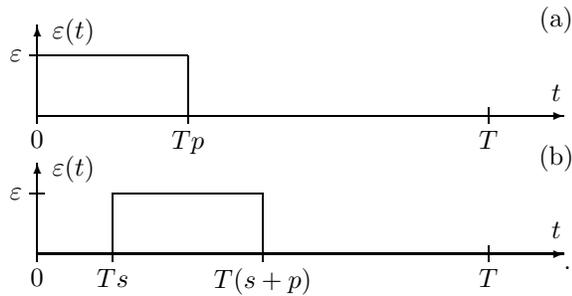}
\caption{Impulse shapes considered for $\varepsilon(t)$,
resulting in the Floquet multipliers
(\ref{eq:floquet_ogy_impulslaenge})
for a finite impulse length (a), and
(\ref{eq:floquet_ogy_impulslaenge_and_s})
if one also adds an additional delay of duration $s$ (b).
\label{schemaPS}
}
\end{figure}

The time-discrete viewpoint now allows to investigate
the influence of timing questions on control.
First we consider the case that the control impulse 
is applied timely in the Poincar\'e section, but 
only for a finite period
$pT$
  within the orbit period ($0<p<1$)
(see Fig.~\ref{schemaPS}a).

\begin{figure}[htbp]
\noindent
\begin{center}
%
\epsfig{file=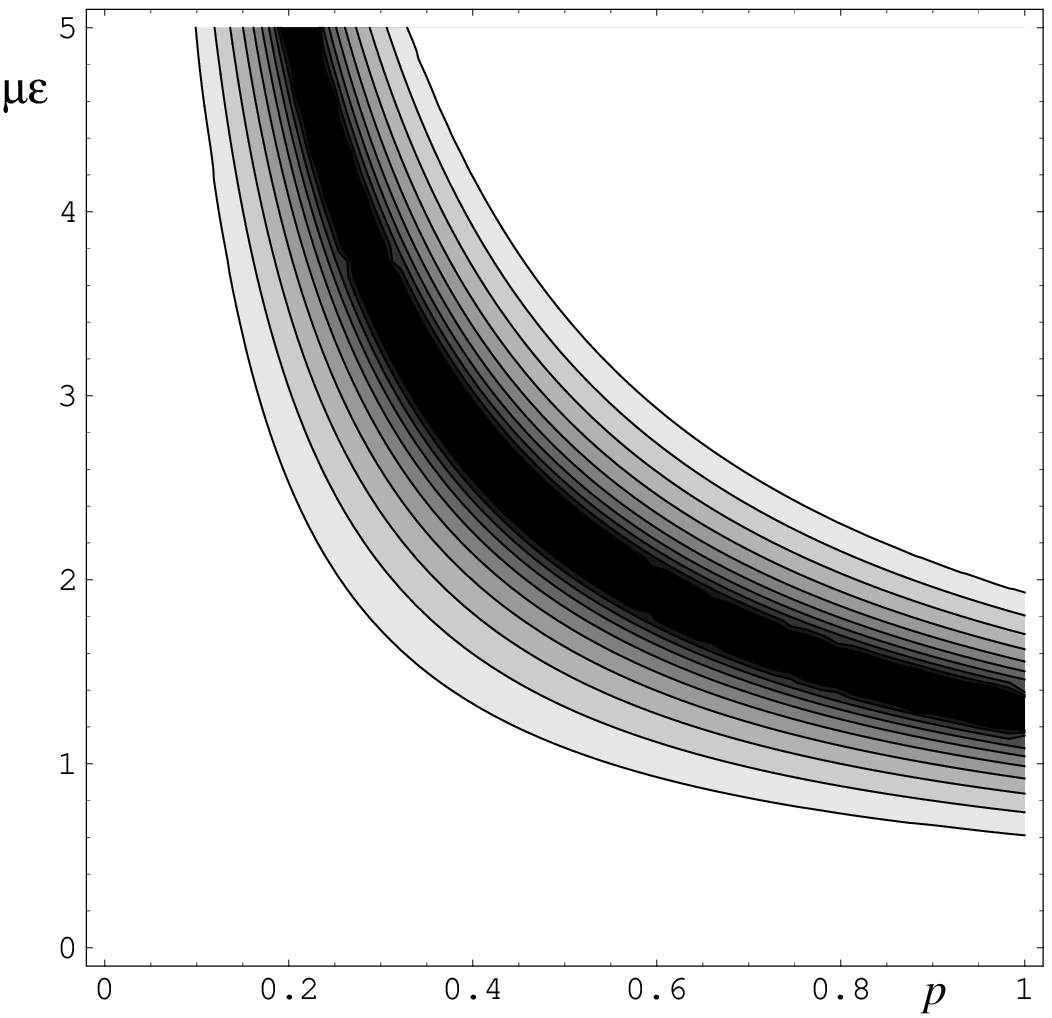,width=0.47\columnwidth} 
%
\raisebox{-1mm}{
\epsfig{file=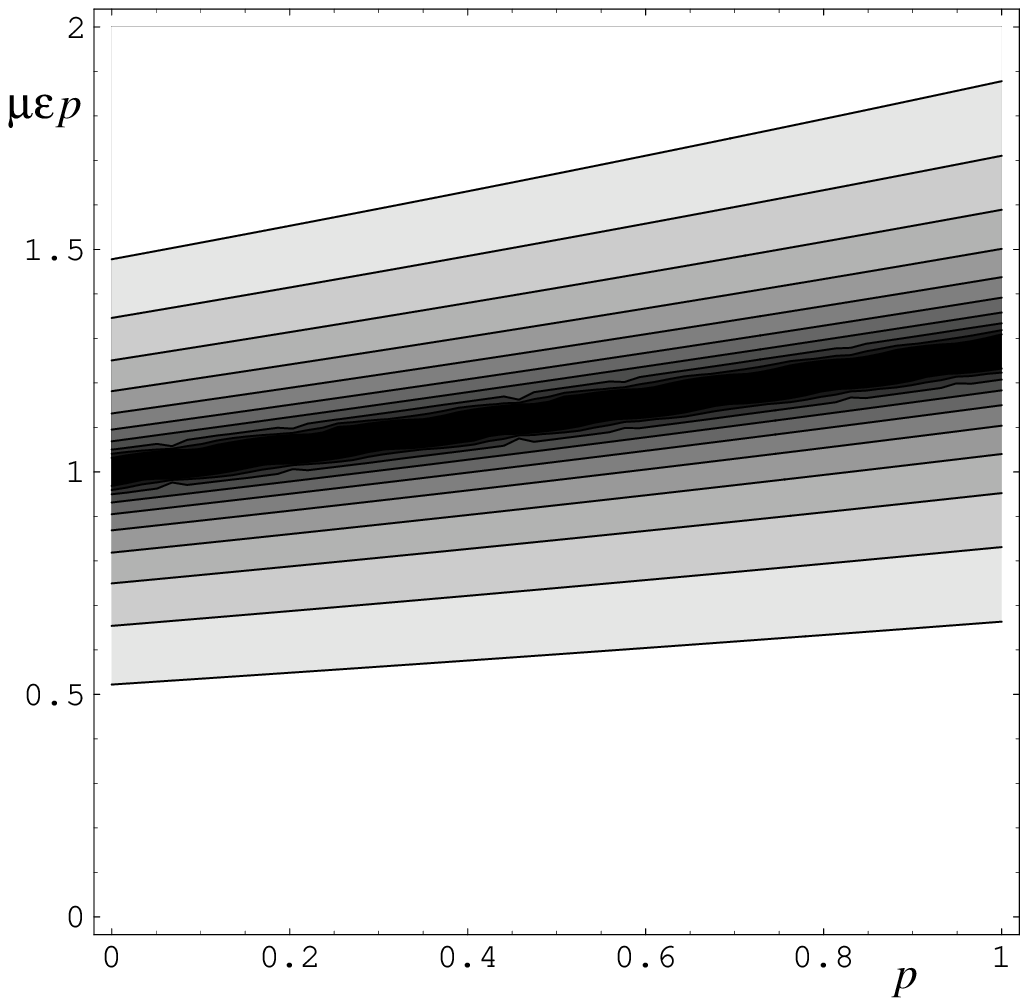,width=0.5\columnwidth} 
}
\end{center}
\caption{Left:
Dependence of OGY control on the duration
 $1\leq pT \leq T$ and strength of control.
Control is possible within the shaded areas,
where darker shadings refer to smaller Re$\gamma$,
including an optimal line inbetween where
Re$\gamma\to{}-\infty$.
In the white areas, Re$\gamma$ is positive,
there the system with applied control becomes unstable.
\label{fig:floq_ogyhyperbel2d} 
Right:
Plot in the  $(p, \mu\varepsilon p)$ plane,
the product $\mu\varepsilon p$
shows only a weak dependency on $p$.
This supports that the linear approximation 
(\ref{eqappro})
is a good approximation 
for (\ref{eq:floquet_ogy_impulslaenge}).
}
\end{figure}

\noindent
This situation is described by the differential equation
\begin{eqnarray}
\dot{x}(t)=\lambda x(t) + \mu\varepsilon x( t-(t ~{\rm{}mod}~ T))
 \Theta((t ~{\rm{}mod}~ T) - p),
\end{eqnarray}
here $\Theta$ 
is a step function  
($\Theta(x)=1$ for $x>0$ and $\Theta(x)=0$ elsewhere).
In the first time interval between $t=0$ and 
$t=pT$
the differential equation reads
$
\dot{x}(t)=\lambda x(t) + \mu\varepsilon x(0).
$
Integration of this  differential equation yields
\begin{eqnarray}
\forall_{0<t \leq pT} \;\;\;\;\;\;\;\;
x(t) 
=
\left((1+  \frac{\mu\varepsilon}{\lambda} )
{\rm{}e}^{\lambda t} - \frac{\mu\varepsilon}{\lambda}
\right) x(0).
\end{eqnarray}
In the second interval
between  $t=pT$  and $t=T$ 
 the differential equation is the same as without control,
$
\dot{x}(t)=\lambda x(t).
$
From this one has immediately
\begin{eqnarray}
\forall_{pT<t<T} \;\;\;\;\;\;\;\;
x(t) = {\rm{}e}^{\lambda (t-pT)}  x(pT)
\end{eqnarray}
\IGN{
If the beginning of the integration period again is denoted by $t$,
this defines an iteration dynamics,
\begin{eqnarray}
x(t+T) &=& 
{\rm{}e}^{\lambda (1-p) T}
\left((1+  \frac{\mu\varepsilon}{\lambda} )
{\rm{}e}^{\lambda pT} - \frac{\mu\varepsilon}{\lambda}
\right)
x(t)
\nonumber
\\
&=&
\left(
\big(1+  \frac{\mu\varepsilon}{\lambda}\big)
{\rm{}e}^{\lambda T} - \frac{\mu\varepsilon}{\lambda}
{\rm{}e}^{\lambda (1-p) T}
\right),
\nonumber
\end{eqnarray} 
}
and the Floquet multiplier of an orbit is
given by
\begin{eqnarray}
{\rm{}e}^{\gamma T}
=
{\rm{}e}^{\lambda T}
\left(
1+\frac{\mu\varepsilon}{\lambda}
(1-{\rm{}e}^{-\lambda pT})
\right).
\label{eq:floquet_ogy_impulslaenge}
\end{eqnarray}
The consequences are shown in 
Fig.~\ref{fig:floq_ogyhyperbel2d}.
One finds that in zero order the
``strength'' 
of control is given by the product
$pT\mu\varepsilon$;
in fact there is a weak linear correction in $p$.
This analysis reproduces well the experimental
results of Mausbach
\cite{mausbachPhd}.
For $\lambda{}pT\leq{}1$ one has
\begin{eqnarray}
{\rm{}e}^{\gamma T}&=& {\rm{}e}^{\lambda T}
(1+\mu\varepsilon pT -
\frac{1}{2} \mu\varepsilon \lambda p^2 T^2 + o(p^3))
\end{eqnarray}
i.e.\ the condition of a constant strength of control reads
\begin{eqnarray}
\mu\varepsilon pT = \frac{1 }{1-\frac{\lambda T}{2}p}
=1 +\frac{\lambda T}{2}p +o(p^2). 
\label{eqappro}
\end{eqnarray}
The result is: Apart from a weak linear correction
for OGY control the length of the impulse can be chosen
arbitrarily,
and the ``strength'' of control in zero order is given
by the time integral over the control impulse.



For the case where the system can only be measured delayed
-- by a delay time of the orbit length and longer --
stability borders \cite{Claussen98b,Just99} 
and improved control schemes have been given in
\cite{Claussen98c,claussenthesis}
 and successfully applied experimentally 
\cite{Claussen98a,mausbach99,klinger01}.
In experimental situations one often has the intermediate 
case that there is a measurement delay $sT$
 that is not neglectable, but within the orbit length.
To keep the case general, we again consider a finite impulse length
$pT$ with
$0<p<1$ und $0<(s+p)<1$,
see
 Fig.~\ref{schemaPS}b.
Again we can integrate piecewise,
and the Floquet multiplier is given by
\begin{eqnarray}
\label{eq:floquet_ogy_impulslaenge_and_s}
{\rm{}e}^{\gamma T}={\rm{}e}^{\lambda T}
\left(
1+\frac{\mu\varepsilon}{\lambda}
{\rm{}e}^{-\lambda sT}
(1-{\rm{}e}^{-\lambda pT})
\right)
\end{eqnarray}
and (\ref{eq:floquet_ogy_impulslaenge})
is included as the special case of $s=0$.

\clearpage
\section{Impulse length influence: Difference control}
Now we analyze the difference control scheme
\cite{Bielawski93a}
\begin{eqnarray}
r(t)=\varepsilon (x(t_{\times})-x(t_{\times\times})), 
\label{eq:difkont_r}
\end{eqnarray}
where $t_{\times}$ and $t_{\times\times}$ denote the times of 
last and last but one Poincar\'e crossing, respectively.
Again the starting point is the linearized equation of motion around
the periodic orbit when control is applied. For difference control
now there is a dependency on two past time steps,
\begin{eqnarray}
\dot{x}(t)&=&\lambda x(t) + \mu\varepsilon x( t-(t ~{\rm{}mod}~ T))
\nonumber\\&&
~~~
- \mu\varepsilon x( t-T-(t ~{\rm{}mod}~ T)).
\label{eq:difkont_kont_dyn}
\end{eqnarray}
Although the right hand side of 
(\ref{eq:difkont_kont_dyn})
depends on $x$ at three different times, it can be nevertheless
integrated exactly, which is mainly due to the fact that
the two past times 
(of the two last Poincar\'e crossings)
have a fixed time difference being equal to
the orbit length.
This allows not only for an exact solution, 
but also offers a correspondence to the time-discrete 
dynamics and the matrix picture used in time-delayed
coordinates.

\begin{figure}[htbp]
\noindent
\centerline{
\epsfig{file=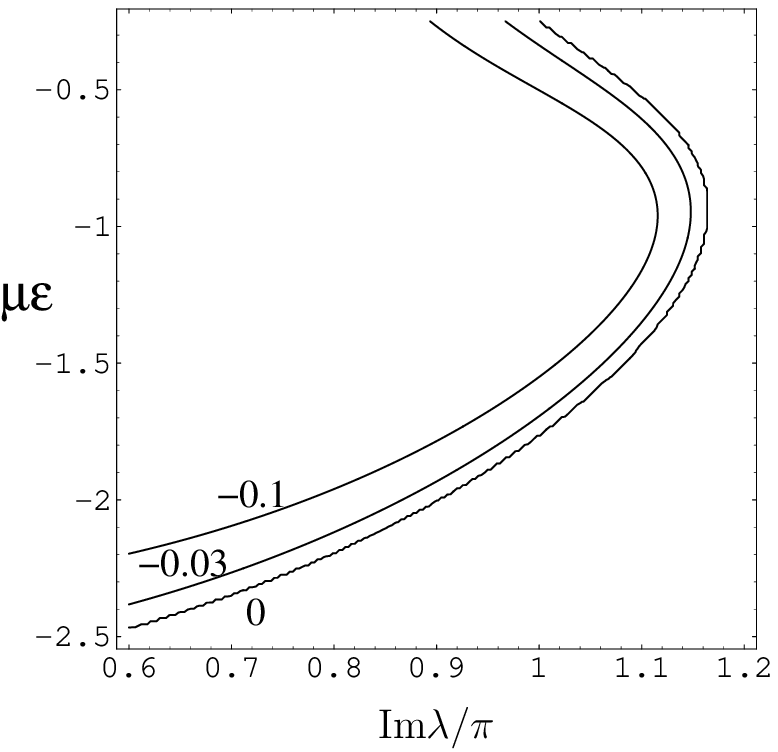,width=0.45\textwidth}
\hspace*{0.08\textwidth}
\epsfig{file=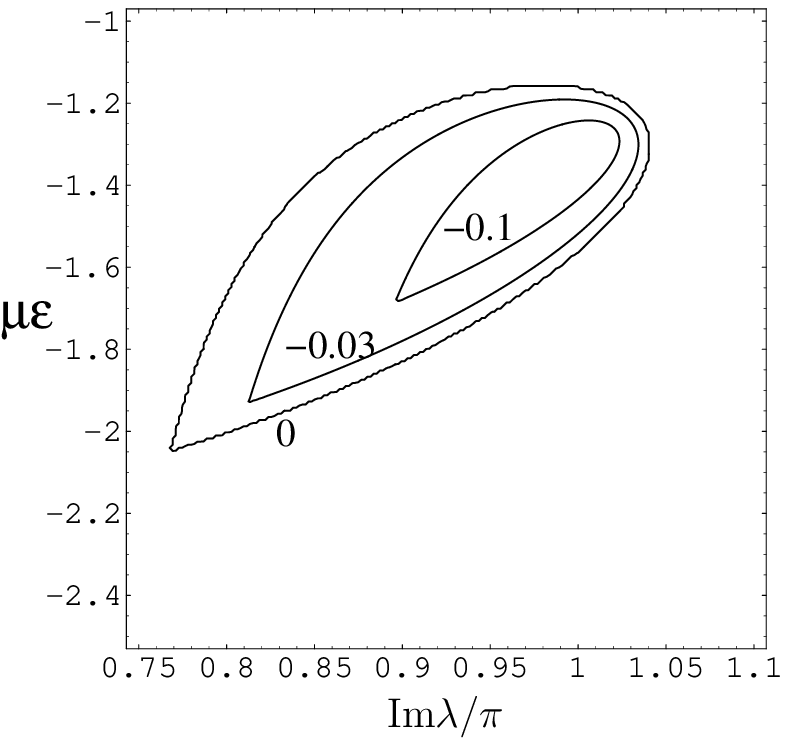,width=0.45\textwidth}
}
\caption{
Time-continuous stability analysis of difference control
with consideration of the length of the control impulse:
Contour plot of the real part of the Floquet exponent in
(Im$\lambda$,$\mu\varepsilon$) space for
$T=1, s=0, p=0.3$,
 Re$\lambda=0.2$ (left) 
and
Re$\lambda=0.6$ (right).
For  ${\rm{}Im}\lambda=\pi$
one finds 
(above Re$\lambda=0.52$) 
an island of stability, which completely disappears for
 $p>p_{\rm max}$ 
 (Fig.\ \ref{fig:floq_diffkont_pmax}).
The Floquet analysis shows that the impulse length is of
fundamental importance for the dynamical behaviour  and
stability of difference control, contrary to
the situation for OGY control.
The contour lines show  the real part of the
Floquet multiplier for
0 (outer), -0.03, and -0.1 (inner contour).
\label{fig:floq_diff_pre.eps} }
\noindent
\centerline{
\epsfig{file=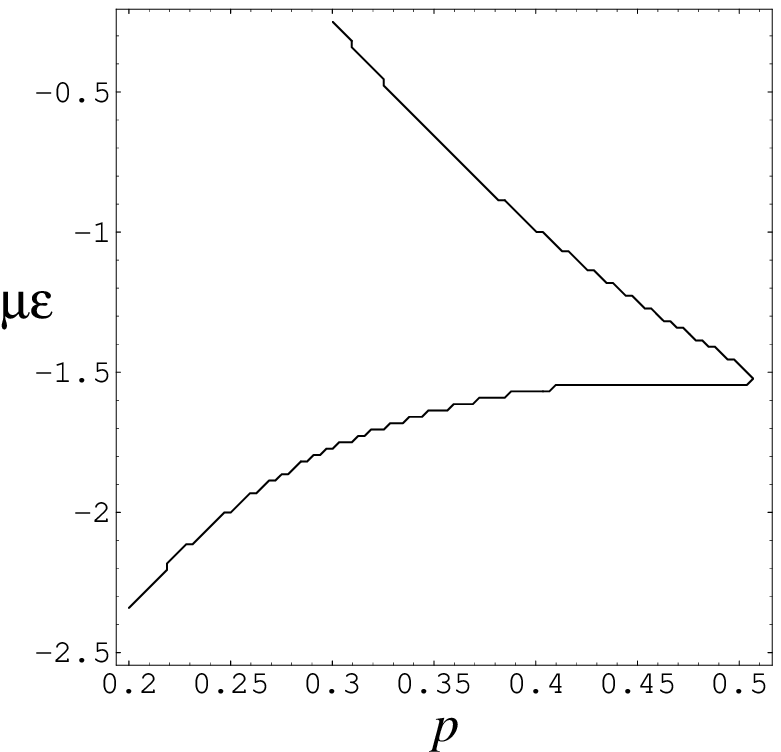,width=0.45\textwidth}
\hspace*{0.08\textwidth}
\epsfig{file=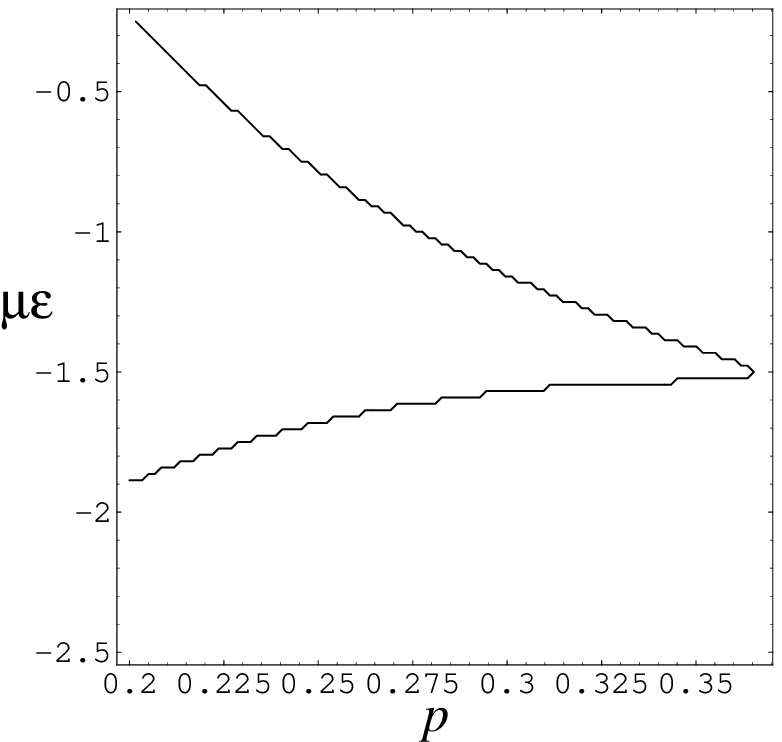,width=0.45\textwidth}
}
\caption{Time-continuous stability analysis of difference control
with consideration of the length of the control impulse:
Stability area for fixed twist Im$\lambda=\pi$ and independence of amplitude $\mu\varepsilon$
and relative impulse length $p$,
for $T=1, s=0$,
Re$\lambda=0.2$ (left) and Re$\lambda=0.6$ (right).
\label{fig:floq_diffkont_keil.eps} }
\end{figure}

\begin{figure}[htbp]
\IGN{
0.000001 0.6663
0.00001 0.6655
0.0001 0.663
0.001 0.657 
0.01 0.637
0.05 0.594
0.1 0.559
0.2 0.505401
0.3 0.462
0.4 0.427
0.5 0.394
0.6 0.367
0.7 0.341
0.8 0.318
0.9 0.296
1.0 0.264
}
\noindent
\centerline{
\epsfig{file=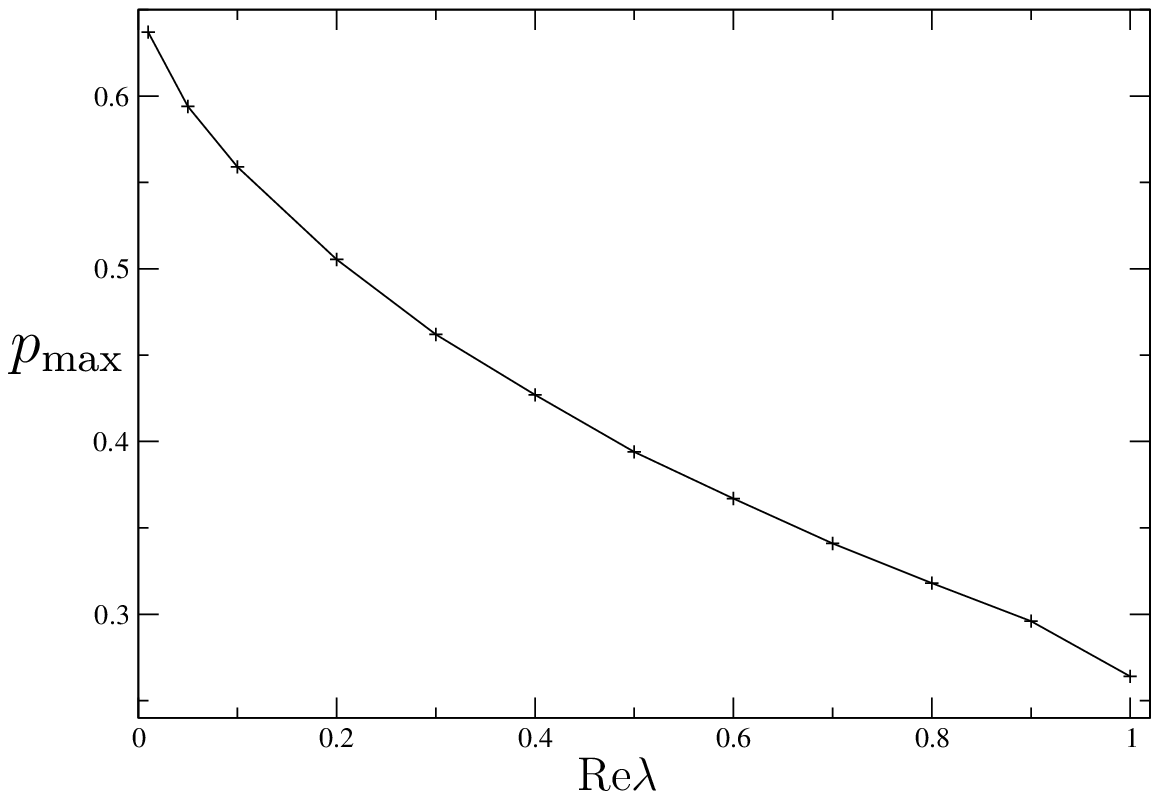,width=0.66\textwidth}
}
\caption{Time-continuous stability analysis of difference control:
Maximal impulse length $p$ for different values of Re$\lambda$ 
in the range from 0.01 to 1.0.
\label{fig:floq_diffkont_pmax} }
\end{figure}


Now also for difference control the experimentally more
common situation of a finite but small measurement delay $sT$
is considered, together with a finite impulse length
$sT$ (here $0<p<1$ and $0<(s+p)<1$).
An analogeous calculation 
\cite{claussenthesis,claussenenoc}
as for the OGY case 
can also be performed 
as follows.
Piecewise integration 
in the first interval $0<t<T\cdot s$,
where the differential equation reads
$\dot{x}(t)=\lambda x(t)$,
yields
again $ x(T\cdot s) = {\rm{}e}^{\lambda T\cdot s}  x(0)$.
In the second interval
$T\cdot s<t<T\cdot (s+p)$
the differential equation reads
\begin{eqnarray}
\dot{x}(t)
= \lambda x(t)
 + \mu\varepsilon (x(0)-x(-T)).
 \nonumber
\end{eqnarray}
Integration yields
\begin{eqnarray}
\nonumber
\forall_{T\cdot s<t<T\cdot (s+p)} \;\;\;\;\;\;\;\;
x(t) &=& -\frac{\mu\varepsilon}{\lambda} (x(0)-x(-T)) \\
&& +\left( \frac{\mu\varepsilon}{\lambda} (x(0)-x(-T))  
+{\rm{}e}^{ \lambda  sT} x(0)
\right) {\rm{}e}^{ \lambda  (t-sT)} 
\nonumber
\\
\nonumber
x(T\cdot (s+p))
 &=&
-\frac{\mu\varepsilon}{\lambda} (x(0)-x(-T)) 
\\
\nonumber
&& 
\frac{\mu\varepsilon}{\lambda} (x(0)-x(-T))
+{\rm{}e}^{ \lambda  pT}
+
{\rm{}e}^{ \lambda  (s+p)T} x(0).
\nonumber
\end{eqnarray}
For the third interval again there is no control, thus
\begin{eqnarray}   
\forall_{T\cdot (s+p)<t<T} \;\;\;\;\;\;\;\;
x(t) =
{\rm{}e}^{\lambda(t-(s+p)T)} x(T\cdot (s+p)).
\nonumber
\end{eqnarray}
\IGN{
<<a0 
<<a3 
--> diffkont_keil.eps

a0:
l=.2
s=0 
T=1 
ll=1.0
Cut[x_] := If[x<0,x,0]

a3:
   
ContourPlot[ Cut[Max[
Re[(1/T) Log[ (1/2)  
* ( Exp[(l + I Pi ll) T]
* (  (1 + (m/(l + I Pi ll)) Exp[- s (l + I Pi ll) T])
* (1 - Exp[- p (l + I Pi ll) T]) ) )
+ (1/2)* Sqrt[ 
( Exp[(l + I Pi ll) T]
* (  (1 + (m/(l + I Pi ll)) Exp[- s (l + I Pi ll) T])
* (1 - Exp[- p (l + I Pi ll) T]) ) )^2
+ 4  ( Exp[(l + I Pi ll) T] (m/(l + I Pi ll))
* Exp[- s (l + I Pi ll) T]
* (1 - Exp[- p (l + I Pi ll) T]) ) ]]],
Re[(1/T) Log[ (1/2)
* ( Exp[(l + I Pi ll) T]
* (  (1 + (m/(l + I Pi ll)) Exp[- s (l + I Pi ll) T])
* (1 - Exp[- p (l + I Pi ll) T]) ) )
- (1/2)* Sqrt[
( Exp[(l + I Pi ll) T]
* (  (1 + (m/(l + I Pi ll)) Exp[- s (l + I Pi ll) T])
* (1 - Exp[- p (l + I Pi ll) T]) ) )^2
+ 4  ( Exp[(l + I Pi ll) T] (m/(l + I Pi ll))
* Exp[- s (l + I Pi ll) T]
* (1 - Exp[- p (l + I Pi ll) T]) )  ]]]
]],{p,0.2,0.5}, {m,-2.5,-.5} ,PlotPoints->100
,PlotRange->{-1.1 , 1}
,Contours ->{ 0}
,ContourShading->False
,ContourSmoothing->Automatic
,FrameLabel->{ "p", FontForm["me",{"Symbol",12}],
 None,None} ,RotateLabel->False
]  

-------------------

floq_diff_insel.eps
generiert aus:


p=.3
l=.5175
l=0.6
s=0 
T=1 
Cut[x_] := If[x<0,x,0]

ContourPlot[ Cut[Max[
Re[(1/T) Log[ (1/2)  
* ( Exp[(l + I Pi ll) T]
* (  (1 + (m/(l + I Pi ll)) Exp[- s (l + I Pi ll) T])
* (1 - Exp[- p (l + I Pi ll) T]) ) )
+ (1/2)* Sqrt[ 
( Exp[(l + I Pi ll) T]
* (  (1 + (m/(l + I Pi ll)) Exp[- s (l + I Pi ll) T])
* (1 - Exp[- p (l + I Pi ll) T]) ) )^2
+ 4  ( Exp[(l + I Pi ll) T] (m/(l + I Pi ll))
* Exp[- s (l + I Pi ll) T]
* (1 - Exp[- p (l + I Pi ll) T]) ) ]]],
Re[(1/T) Log[ (1/2)
* ( Exp[(l + I Pi ll) T]
* (  (1 + (m/(l + I Pi ll)) Exp[- s (l + I Pi ll) T])
* (1 - Exp[- p (l + I Pi ll) T]) ) )
- (1/2)* Sqrt[
( Exp[(l + I Pi ll) T]
* (  (1 + (m/(l + I Pi ll)) Exp[- s (l + I Pi ll) T])
* (1 - Exp[- p (l + I Pi ll) T]) ) )^2
+ 4  ( Exp[(l + I Pi ll) T] (m/(l + I Pi ll))
* Exp[- s (l + I Pi ll) T]
* (1 - Exp[- p (l + I Pi ll) T]) )  ]]]
]],{ll,0.75,1.1}, {m,-2.5,-1} ,PlotPoints->200
,PlotRange->{-1.1 , 1}
,Contours ->{ -0.000001, -0.03, -.1}
,ContourShading->False
,ContourSmoothing->400
,FrameLabel->{ "Im l/p", FontForm["me",{"Symbol",12}],
 None,None} ,RotateLabel->False
]

l=0.6
{ll, 0.75, 1.1}, {m, -2.5, -1.0}

l=0.2
{ll, 0.6, 1.2}, {m, -2.5, -0.25}
FrameLabel -> {"", "", None, None}, RotateLabel -> False]

Keil:

l=0.6
{p, 0.2, 0.367}, {m, -2.5, -.25}

l=0.2
{p, 0.2, 0.51}, {m, -2.5, -.25}

While the previous figures were generated on a 30 MHz
Sparc station with a running time above 1 hour each,
I took the opportunity to perform additional parameter
explorations that were practically impossible at that time.
}
%
%
%
%
%
%
%
%
%
%
\noindent
Collecting together, we find for $x(T)$ 
\begin{eqnarray}
\nonumber
x(T)&=&x(0){\rm{}e}^{\lambda T}
\left(
1+\frac{\mu\varepsilon}{\lambda}
{\rm{}e}^{-\lambda sT}
(1-{\rm{}e}^{-\lambda pT})
\right)
\nonumber   
\\ 
&& -x(-T) {\rm{}e}^{\lambda T}
\frac{\mu\varepsilon}{\lambda}
{\rm{}e}^{-\lambda sT}
(1-{\rm{}e}^{-\lambda pT})
\end{eqnarray}
or, in time-delayed coordinates of the last and last but one 
Poincar\'e crossing
\small
\begin{eqnarray}    
\left(\!\!\begin{array}{c}x_{n+1}\\x_{n}\end{array}\!\!\right)
\!=\! 
\nonumber
\mbox{\small$
\left(
\!\!\!
\begin{array}{cc}
{\rm{}e}^{\lambda T}
\! \left(
\!
1
\!+
\!
\frac{\mu\varepsilon
(1-{\rm{}e}^{-\lambda pT})
}{\lambda
{\rm{}e}^{\lambda sT}
}
\!\right)
\!
& -{\rm{}e}^{\lambda T}
\frac{\mu\varepsilon
(1-{\rm{}e}^{-\lambda pT})
}{\lambda
{\rm{}e}^{\lambda sT}
}
\\
1 & 0
\end{array}
\!\!
\right)
$}
\!\!
\left(\!\!\begin{array}{c}x_{n}\\x_{n-1}\end{array}\!\!\right)\!\!.
\nonumber 
\end{eqnarray}
\normalsize
\noindent
If we identify with the coefficients of the time-discrete case,
\mbox{$\lambda_{\sf d}={\rm{}e}^{\lambda T}$} and
 $\mu_{\sf d} \varepsilon_{\sf d} 
={\rm{}e}^{-\lambda s T} (1- {\rm{}e}^{\lambda p  T})
 \frac{\mu\varepsilon}{\lambda} $,
the dyna\-mics in the 
 Poincar\'e iteration $t=nT$
becomes identical with the pure discrete description;
this again illustrates the power of the concept of the
Poincar\'e map.
In principle, 
 due to the low degree of the characteristic polynomial, 
one could explicitely diagonalize the 
iteration matrix, allowing for a closed expression for the
$n$-th power of the iteration matrix. 
However, for the stability analysis only the eigenvalues
are needed.
%
%
%
%
%
 For the Floquet multiplier one has
\begin{eqnarray}
\nonumber
{\rm{}e}^{2\gamma T}&=&
{\rm{}e}^{\gamma T}
{\rm{}e}^{\lambda T}
\left(
1+\frac{\mu\varepsilon}{\lambda}
{\rm{}e}^{-\lambda sT}
(1-{\rm{}e}^{-\lambda pT})
\right)
\\ 
&&
- {\rm{}e}^{\lambda T}
\frac{\mu\varepsilon}{\lambda}
{\rm{}e}^{-\lambda sT}
(1-{\rm{}e}^{-\lambda pT}).
\end{eqnarray}
This quadratic equation
yields two Floquet multipliers,
\small
\begin{eqnarray}
\nonumber
{\rm{}e}^{\gamma T} &=& 
 \frac{1}{2}
{\rm{}e}^{\lambda T} 
\left(
1+\frac{\mu\varepsilon}{\lambda}
{\rm{}e}^{-\lambda sT}
(1-{\rm{}e}^{-\lambda pT})
\right)
\\&& \nonumber
\pm\frac{1}{2} \sqrt{
 \left({\rm{}e}^{\lambda T} 
\big(
1+\frac{\mu\varepsilon}{\lambda}
{\rm{}e}^{-\lambda sT}
(1-{\rm{}e}^{-\lambda pT})
\big)
\right)^2
~~~~~~~} \nonumber\\
&&
\mbox{} ~~~~~~~~~~~~~~~~~~~~~~~~~~~
\overline{~~~~~~~
+ 4  {\rm{}e}^{\lambda T}
\frac{\mu\varepsilon}{\lambda}
{\rm{}e}^{-\lambda sT}
(1-{\rm{}e}^{-\lambda pT})
}.
\nonumber
\end{eqnarray}
\normalsize
%
Figures \ref{fig:floq_diff_pre.eps}
and \ref{fig:floq_diffkont_keil.eps}
show, by example of  Re$\lambda=0.2$,
(resp.\ Re$\lambda=0.6$)
that at  Im$\lambda=\pi$  for $p<p_{\rm max}({\rm Re}\lambda=0.2)\simeq 0.505401$ 
(resp $p<p_{\rm max}({\rm Re}\lambda=0.6)\simeq 0.367$)
 there exists an island of stability,
whose width decreases to zero for $p\to p_{\rm max}$.
Here explicitely 
the influence of
the impulse length
can be seen. 
The maximal value of $p_{\rm max}$ is 
shown in Fig.~\ref{fig:floq_diffkont_pmax}.
%


%
\clearpage


%
\section{Discussion of the time-continuous model: Relaxing the assumption of a constant local Ljapunov exponent}
\begin{figure}[hbtp]
\epsfig{file=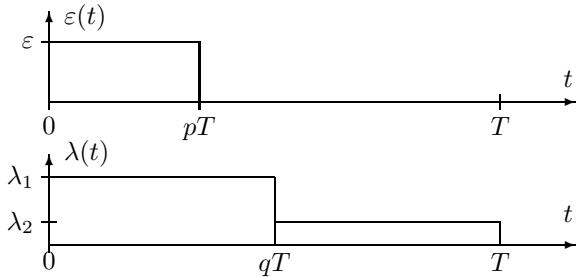}
\caption{Impulse shapes  $\varepsilon(t)$
and time-varying $\lambda(t)$:
Case of a short impulse $p<q$.
\label{schemaEL}
}
\end{figure}

The quantitative analysis given above was based on the
model assumption that one has a constant local Ljapunov exponent
around the orbit -- a condition that will almost never be
fulfilled exactly for a typical orbit of a chaotic system.
To test whether this assumption is crucial, 
the stability 
 can be investigated for OGY control by assuming $\lambda$ to
be time-dependent along the orbit. 
For simplicity, we consider the exemplaric case
shown 
in Fig.~\ref{schemaEL}
 that 
$\lambda(t)=\lambda_1$ for $0<t < qT$ and 
$\lambda(t)=\lambda_2$ for  $qT < t < T$,
generalizing the case $q=1/2$ already sketched in
\cite{claussenenoc}.

\subsection{The case $p<q$ of a short impulse}
\noindent
If we assume $p<q$, we have for the OGY case,
\begin{eqnarray}
\nonumber
x(pT)&=& 
( {\rm{}e}^{\lambda_1 pT} 
(1+  (\mu\varepsilon / \lambda_1))
- (\mu\varepsilon / \lambda_1) )
x(0)
\\
x(qT)&=& {\rm{}e}^{\lambda_1 (qT - pT)} x(pT)
\nonumber
\\
x(T)&=& {\rm{}e}^{\lambda_2 (T-qT)} x(qT).
\end{eqnarray}
Using
$\bar{\lambda}:=q\lambda_1+(1-q)\lambda_2$,
the Floquet multiplier reads now
\begin{eqnarray}
{\rm{}e}^{\gamma T}
=
{\rm{}e}^{\bar{\lambda} T}
\left(
1+\frac{\mu\varepsilon}{\lambda_1}
(1-{\rm{}e}^{-\lambda_1 pT})
\right)
\label{eq:floquuet_ogy_lambda12}
\end{eqnarray}
in contrast to eq.~(\ref{eq:floquet_ogy_impulslaenge}).
%
Again in zero order the ``strength'' of control is given by the 
product $p\mu\varepsilon$;
in first order 
$\lambda_1{}pT\leq{}1$
again the weak linear dependence on $p$
applies,
\begin{eqnarray}
{\rm{}e}^{\gamma T}
= 
{\rm{}e}^{\bar{\lambda} T}
(1+\mu\varepsilon pT (1-\frac{1}{2} \lambda_1 p T + o(p^2) )),
\end{eqnarray}
i.e.\
 for
a constant ``strength'' of control,
one has to fulfill
\begin{eqnarray}
\mu\varepsilon pT = \frac{1 }{1-\frac{\lambda_1 T}{2}p}
+o(p^2)
=1 +\frac{\lambda_1 T}{2}p +o(p^2). 
\end{eqnarray}
Thus, the value of $\lambda_1$, i.e.\ here,
the deviation of $\lambda(t)$ from its average
value  $\bar{\lambda}$ during the control impulse,
only contributes in first order.
More complicated cases can be tackled likewise, 
giving corrections for the quantitative 
$p$-dependence of the optimal control gain $\varepsilon$,
but preserving the qualitative behaviour 
discussed above.

\subsection{The case $p>q$ of a long impulse}
\begin{figure}[hbtp]
\epsfig{file=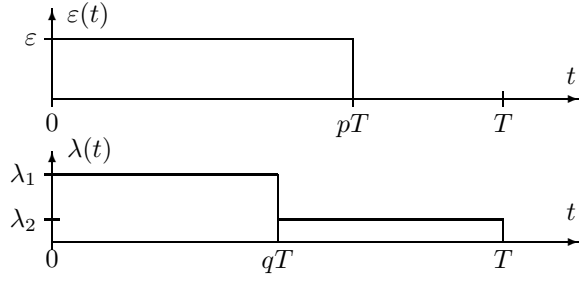}
\caption{Impulse shapes  $\varepsilon(t)$
and time-varying $\lambda(t)$:
Case of a long impulse $p>q$.
\label{schemaELlong}
}
\end{figure}

While a short impulse is the experimentally more 
feasible case, for completeness, also the $p>q$ shown in
Fig.~\ref{schemaELlong}
can be investigated in this manner.
Here we have two time intervals where the control is 
active; and in the second one the initial condition
$x(pT)$ has to be distinguished from the
position $x(0)$ at the last Poincar\'e crossing,
from which the control value is calculated and which
determines the inhomogenity of the ODE. 

\noindent
Integration over the three time intervals yields
\begin{eqnarray}
x(qT)
 &=& 
 {\rm{}e}^{\lambda_1 q T} x(0) \Big(1+\frac{\mu\varepsilon}{\lambda_1}\Big)
 -  x(0) \frac{\mu\varepsilon}{\lambda_1}
\nonumber
\\
x(pT) &=&
 {\rm{}e}^{\lambda_2 (pT-qT)}
\Big(x(qT)+x(0)\frac{\mu\varepsilon}{\lambda_2}\Big)
-x(0)\frac{\mu\varepsilon}{\lambda_2}
\nonumber
\\
x(T) &=&  {\rm{}e}^{\lambda_2 (T-pT)}  x(pT)
\nonumber
\end{eqnarray}
so that we arrive at
\begin{eqnarray}
x(T)
 &=& x(0) {\rm{}e}^{\bar{\lambda}T}
\Bigg[  1+ \frac{\mu\varepsilon}{\lambda_1} 
(1-{\rm{}e}^{-\lambda_1 q T})
+ \frac{\mu\varepsilon}{\lambda_2} 
(1-{\rm{}e}^{-\lambda_2 (p-q) T}) {\rm{}e}^{-\lambda_1 q T}
\Bigg].
\end{eqnarray}

\vspace{12mm}

\subsubsection*{Weakly nonlinear approximation for $q=1/2$}
As $p>q$, and $q$ is a fixed value for  the given system, 
a discussion of 
$p \lambda T \ll 1$ 
can no longer be based on the $p \to 0$ case.
As $p$ is of order 1, 
an expansion as above is meaningful only for
the case where $\lambda T \ll 1$,
i.e., we derive an approximation for those
UPOs which have an only marginally positive 
Floquet multiplier.
For $q=1/2$ we now explicitely discuss this
``weakly nonlinear'' case $\lambda_1 T \ll 1$, $\lambda_2 T \ll 1$,
\begin{eqnarray}
\nonumber
\!\!\!\!\!\!\!\!\!\!\!\!\!\!\!\!\!\!\!\!\!
x(T) &=& x(0) {\rm{}e}^{\bar{\lambda}T}
\Bigg[ 1+ \frac{\mu\varepsilon}{\lambda_1} \Big( -\lambda_1
\frac{T}{2} +\frac{\lambda_1^2 T^2}{8} - \frac{\lambda_1^3 T^3}{48}\Big)
\\
\nonumber
&& ~~~~~~~~~~~~~
+\frac{\mu\varepsilon}{\lambda_2} \Big(-\lambda_2 T (p-\frac{1}{2}) +
\frac{\lambda_2^2T^2(p-\frac{1}{2})^2}{2}\Big) \Big(-\frac{\lambda_1 T}{2} +
\frac{\lambda_1^2 T^2}{8} \Big) \Bigg]
\\
\!\!\!\!\!\!\!\!\!\!\!\!\!\!\!\!\!\!\!\!\!
\nonumber
&=&
x(0) {\rm{}e}^{\bar{\lambda}T}
\left[1-\frac{\mu\varepsilon T}{2} \Big(1-\lambda_1 T (p-\frac{1}{4}) +
o(\lambda_1^2,\lambda_1 \lambda_2, \lambda_2^2) \Big) \right]
\nonumber
\end{eqnarray}
that is, control is kept constant in lowest order for
\begin{eqnarray}
\frac{\mu\varepsilon T}{2}  \stackrel{!}{=} \frac{1}{1-\lambda_1 T
(p-\frac{1}{4})}
\simeq 1 + \lambda_1 T(p-\frac{1}{4})
\end{eqnarray}
for orbits sharing the same value of $\bar{\lambda}$.

\subsection{Difference control and nonconstant local Ljapunov exponent}
For completeness, now also the case of difference control
is considered. As has been shown before, for the case of a constant
Ljapunov exponent, impulse lengths of $p>\frac{1}{2}$ 
do not lead to stable control; therefore 
the case $p>q$ is completely irrelevant, and
only the case
$p<q$ has to be be considered.
In the first interval, $\lambda_1$ is active and integration yields
\begin{eqnarray}
x(pT)={\rm{}e}^{\lambda_1 pT}
\left[ x(0) \big( 1+\frac{\mu\varepsilon}{\lambda_1} ( {\rm{}e}^{-\lambda_1 pT}-1) \big)
-x(-T) \frac{\mu\varepsilon}{\lambda_1} ( {\rm{}e}^{-\lambda_1 pT}-1) 
\right]
\end{eqnarray}
and the subsequent intervals have the control switched off,
\begin{eqnarray}
x(T)&=& 
{\rm{}e}^{\lambda_2 (1-q)T} x(qT)
=
{\rm{}e}^{\lambda_2 (1-q)T}
{\rm{}e}^{\lambda_1 (q-p)T}
x(pT)
\\
&=& {\rm{}e}^{\bar{\lambda}T} 
\left[ x(0) \big( 1+\frac{\mu\varepsilon}{\lambda_1} ( {\rm{}e}^{-\lambda_1
pT}-1) \big)
-x(-T) \frac{\mu\varepsilon}{\lambda_1} ( {\rm{}e}^{-\lambda_1 pT}-1) 
\right].
\end{eqnarray}
Again we use the average value
$\bar{\lambda}=\lambda_1 q + \lambda_2 (1-q)$ 
to simplify the expressions,
and
 the coordinates in the Poincar\'e countings $x_{n+1}=x(T)$, 
$x_n=x(0)$, $x_{n-1}=x(-T)$.
 We have
\begin{eqnarray}
\left(\begin{array}{c}x_{n+1} \\ x_n \end{array}\right) =
\left(\begin{array}{cc}
{\rm{}e}^{\bar{\lambda}T} 
(1+\frac{\mu\varepsilon}{\lambda_1}( {\rm{}e}^{-\lambda_1 pT}-1))
&
-{\rm{}e}^{\bar{\lambda}T}
\frac{\mu\varepsilon}{\lambda_1}( {\rm{}e}^{-\lambda_1 pT}-1)
\\ 1&0 \end{array}\right)
\left(\begin{array}{c}x_{n} \\ x_{n-1} \end{array}\right) 
\nonumber\\
\end{eqnarray}
which leads to the characteristic equation for the two
Floquet multipliers ${\rm{}e}^{\gamma T}$
\begin{eqnarray}
{\rm{}e}^{2 \gamma T} =
{\rm{}e}^{\gamma T}   
 {\rm{}e}^{\bar{\lambda}T}
\Big(1+\frac{\mu\varepsilon}{\lambda_1}( {\rm{}e}^{-\lambda_1 pT}-1)\Big)
- {\rm{}e}^{\bar{\lambda}T}
\frac{\mu\varepsilon}{\lambda_1}( {\rm{}e}^{-\lambda_1 pT}-1).
\end{eqnarray}
This generalizes the discussion of difference control to the
case of nonconstant $\lambda$.

\section{Conclusions and Outlook}
To summarize, a new time-continuous stability analysis 
of Poincar\'e-based control methods was introduced.
This general and novel approach allows to investigate timing questions 
of Poincar{\'e} based control schemes that cannot be
analyzed within the picture of the Poincar\'e{} iteration.
For both OGY and difference control
it has been possible for a homogeneous case to
integrate the dynamics exactly.
While for OGY control the impulse length turns out not
to be crucial, for difference control it is, and the
impulse has to be shorter than a critical
fraction of the period, which is
of the order of half of the period
and decreases for larger Ljapunov exponents.
Such timing dependence is not completely uncommon in
feedback systems with delay;
in time-continuous 
feeedback control, a half-period feedback resulted 
in an enlarged stability range  \cite{choe07}.

Techniques of chaos control, be it Poincar\'e-based, following
the Pyragas technique, or open-loop
\cite{claussenPOR,lima90,filatrella93,mettin,kivshar}
have been of great interest not only in technical
systems,
but also in biologically, especially neural systems
and excitable media
\cite{balanov06,tass}.
Besides the approach of controlling pathological neural subsystems directly,
the intact brain already bears implementations of feedback control
\cite{dahlem08},
implying that the failure of the respective circuits eventually 
results in migraine or stroke.
Also the human gait system, as virtually any 
perception-motor system, 
performs control of a bio-mechanical system
with delays; 
and disturbances of the delay loops as well
as the cortical control may result in
tremor and related movement disorders 
\cite{gait,tremor,govindan05,timmer}.
In the thalamocortical system, a designated impulse shape,
formed by the so-called slow waves that emerge in the
cortex during S2 sleep,
has been shown to act as an open-loop controller of 
thalamic oscillator networks \cite{mayer07}.
This offers further possibilities to
influence human sleep in the case of sleep disturbances:
Recent control techniques by transcranial electrical 
or magnetic stimulation 
\cite{tcs,stickgold,siebner04,siebner05}
have been demonstrated to influence human sleep as
well as to affect memory consolidation during sleep.
In most of these techniques, the impulse shape and
relative duration of 
the control impulse has significant impact on the results,
thus different control goals may become 
accessible within the same setup.
For the systematic understanding how such control techniques
influence the brain, 
detailed models are of likewise importance as 
the methodical understanding of 
the theoretically possible control methods.

 \subsection*{Acknowledgments} The author acknowledges financial support from Deutsche Forschungsgemeinschaft (DFG) through SFB-654 ``Plasticity and Sleep''.

\end{document}